# Overlooked weak structural connections support human cognition under nonlinear connectome scaling

Rong Wang[1#,*], Zhao Chang[2,3#], Xuechun Liu[1], Daniel Kristanto[4], Étienne Gérard Guy Gartner[2,3], Xinyang Liu[5], Mianxin Liu[6], Ying Wu[1], Ming Lui[7], Changsong Zhou[2,3,,8,9*]

1. State Key Laboratory for Strength and Vibration of Mechanical Structures, School of Aerospace Engineering, Xi'an Jiaotong University, Xi'an 710049, China
2. Department of Physics, Hong Kong Baptist University, Kowloon Tong, Hong Kong
3. Centre for Nonlinear Studies and Beijing-Hong Kong-Singapore Joint Centre for Nonlinear and Complex Systems (Hong Kong), Institute of Computational and Theoretical Studies, Hong Kong Baptist University, Kowloon Tong, Hong Kong
4. Department of Psychology, Carl von Ossietzky Universität Oldenburg, Oldenburg, 26129, Germany
5. Department of Radiology and Biomedical Imaging, University of California San Francisco, San Francisco, California, USA
6. Shanghai Artificial Intelligence Laboratory, Shanghai, 200232, China
7. School of Psychology, University of Roehampton, London
8. Life Science Imaging Centre, Hong Kong Baptist University, Kowloon Tong, Hong Kong;
9. Department of Physics, Zhejiang University, 38 Zheda Road, Hangzhou, 310000, China

[#]Rong Wang and Zhao Chang contributed equally to this work.
*Corresponding authors: tobous@mail.xjtu.edu.cn, cszhou@hkbu.edu.hk

## ABSTRACT

Human cognition depends on large-scale communication constrained by white-matter architecture. Although weak connections are abundant in mammalian connectomes, they have long been treated as noise and downweighted because of tractography uncertainty in the human brain, and their relevance to human cognition and large-scale functional organization remains unresolved. Across multiple datasets and tractography pipelines, we show that, when tractography-derived connection weights are interpreted through a nonlinear weighting framework, weak connections make measurable contributions to cognitive predictions, functional-connectivity simulation, and structure–function coupling. These effects are selective: nonlinear weighting improves the prediction of general cognitive ability and memory more than that of crystallized intelligence or processing speed, consistent with the notion that weak connections preferentially expand the modal repertoire of brain structural networks to enhance both large-scale integration and fine-grained segregation, thereby supporting the functional balance essential for diverse cognitive abilities. Importantly, these effects are replicated in a reliability-aware connectome generated by integrating two post-tractography filtering methods, in which preserving weak links consistently outperforms conventional thresholding strategy in structure-function coupling and model-based simulation. Finally, we show that weak connections contain functional information organized along system-level and gene-expression gradients. In particular, weak connections with negative gene co-expression predominantly link visual/motor with limbic regions and exert a disproportionately large influence on brain function than those with positive gene co-expression. Our findings provide compelling evidence that weak connectivity plays an underestimated yet essential role in human cognition under nonlinear scaling and establish a refined framework for reliably characterizing brain structural connectivity.

## Introduction

Human cognition emerges from large-scale interactions constrained by the brain structural connectome, in which cortical and subcortical regions are linked through white-matter pathways [1, 2]. Rather than being well described as sparse binary graphs, brain structural connectomes exhibit heavy-tailed weight distributions spanning several orders of magnitude, with numerous weak connections embedded alongside the strong pathways. Diffusion MRI tractography provides the only noninvasive approach for reconstructing these pathways in vivo, commonly referred to as structural connectivity (SC) [3, 4]. Despite well-known limitations [3-6], including false positives and uncertainty in mapping white-matter pathways, tractography has substantially advanced the study of brain function, cognition, and neuropsychiatric disorders [4, 7-10], and serves as a foundational input for large-scale brain modeling platforms such as The Virtual Brain [11] and Digital Twin Brain [12]. However, the functional relevance of the widespread weak connectivity within these dense connectomes remains poorly understood.

A major challenge is that weak connectivity occupies an ambiguous position between biological organization and tractography uncertainty [13-15]. Diffusion MRI–derived weak connections are also the most vulnerable to false positives, distance-dependent attenuation, and crossing-fiber uncertainty [3, 14, 15]. Consequently, many studies downweighted, filtered, or thresholded low-weight streamlines using approaches including thresholding, SIFT2 [16] and COMMIT2 [3], resulting in sparser and more reliable connectomes [17-23]. At the same time, removing a large proportion of the weakest connections (e.g., 50–90%) indeed had little effect on graph-theoretical properties of brain networks [17], and removing the ~50% weakest connections had negligible effects on the neural signal-propagation ratio in a rat-brain dynamic model [24]. However, invasive tract-tracing studies across mammalian brains consistently reveal dense cortical architectures with connection weight spanning several orders of magnitude (e.g., 97% for 47 areas in mouse brains) [25-28]. These findings indicate that connection weight is better viewed as a continuum without a universally accepted cutoff, and weak connectivity is a natural and pervasive feature of biological networks [13, 27, 29]. Therefore, the critical question is not only whether tractography-derived weak connections are reliable, but under what weighting assumptions they can become functionally relevant to brain function and cognition.

A common implicit assumption in network neuroscience is that the functional influence of structural connectivity scales linearly with tractography weight. Under this assumption, weak connections may exert negligible influence on graph metrics, brain dynamics, and cognitive predictions [17, 24] because their weights are several orders of magnitude smaller than those of strong connectivity. Instead of imposing an arbitrary threshold, we therefore hypothesized that the effective functional influence of structural connectivity may follow a nonlinear scaling framework rather than a strictly linear weighting rule. To test such a possibility, we used a phenomenological nonlinear scaling model in which connection weight $w_{ij}$ was transformed as $w_{ij}^{\beta}$. Here, $\beta$ is an effective, target-dependent weighting parameter rather than a universal biological constant. When $\beta$ = 1.0, the model reduces to the conventional linear assumption in previous studies [17-23]; when 0.0< $\beta$ < 1.0, the transformation selectively amplifies the effective influence of weak connectivity relative to strong connectivity, but also reduces the dynamic range of connection weights. Under this nonlinear weighting framework, weak connections could make disproportionately large contributions to brain functional organization and cognition despite their small weights.

Here, using the HCP (*n*=999) and UCLA datasets (*n* = 122), we constructed SC matrices with multiple tractography and filtering pipelines. First, we constructed a connectome-based machine learning model to investigate whether nonlinear weighting reveals the functional relevance of weak connectivity to cognitive abilities and whether the functional relevance persists after accounting for likely tractography false positives. Then, in a large-scale brain dynamic model, we studied how weak connectivity supports brain functional organization under nonlinear weighting. We further performed an eigenmode analysis without thresholding. This analysis examined how nonlinear weighting alters the contribution of weak connectivity across hierarchical activation modes. It provides a potential mechanistic explanation for the role of weak connectivity in cognitive abilities. Finally, we examined the organizational principle of weak connections in the brain in terms of functional systems and gene co-expression, and investigated the heterogeneous functions of weak connectivity in relation to brain function and cognitive abilities.

## Results

### Nonlinear weighting reveals functional relevance of weak connectivity to cognitive abilities

We used probabilistic tractography [30] to reconstruct streamlines and construct whole-cortex basic SC networks (*N*=360 regions, binary density: 56.3±3.8%). SC weights span several orders of magnitude across individuals (**Fig. S1**), consistent with observations of other mammalian brains using reliable retrograde tracer injections [25-27, 31]. We tested the phenomenological model in which the effective influence of SC scales nonlinearly with connection weight such that weak connections may exert more functional influence than their raw weights suggest. We first extracted individual estimates of crystallized intelligence (***cry)***, processing speed (***spd)***, general cognitive ability (***g)*** and memory (***mem***) through a structural equation model (SEM) (**Fig. S2A**). These four latent abilities were selected to probe different functional demands on network segregation (required by crystallized intelligence and processing speed), integration (required by general cognitive ability) and their balance (required by memory) [32]. We then used a connectome-based prediction model (CPM) [33] to predict cognitive abilities with scaled connection weight $w_{ij}^{\beta}$ (**Fig. S2B**). We found that general cognitive ability was predicted above chance from SC, memory and crystallized intelligence shown only modest predictive associations (**Fig. 1A**), and processing speed cannot be predicted (**Fig. S3**). More importantly, $\beta < 1.0$ appeared to generate better predictions. To test this improvement while avoiding optimistic bias, we introduced an inner loop to select the optimal $\beta$ inside a fully nested cross-validation in the CPM (see Materials and Methods). The optimal $\beta$ values were approximately $\beta = 0.703$ $(95\%\ \mathrm{CI}, 0.641 - 0.765)$ for general cognitive ability, and $\beta = 0.602$ $(95\%\ \mathrm{CI}, 0.532 - 0.672)$ for memory, and the selection of optimal $\beta$ was relative stable across 100 cross-validation splits (**Fig. S4**). More importantly, the predictions for optimal $\beta$ were higher than those for $\beta = 1.0$ (***g***: paired t-test, $t_{pt}(99) = 7.344$, $p_{pt} < 0.001$, Nadeau-Bengio correction, $t_{NB}(99) = 2.110$, $p_{NB} = 0.037$; ***mem***: $t_{pt}(99) = 7.549$, $p_{pt} < 0.001$, $t_{NB}(99) = 2.169$, $p_{NB} = 0.032$). We further used rank correlation to ensure the independence of feature selection on $\beta$, such that $\beta$ only affects the prediction procedure (Eq. 2). We also found that $\beta$<1.0 still generated better predictions than $\beta$=1.0 for general cognitive ability and memory (**see Fig. S5**). For crystallized intelligence, the optimal $\beta$ was $0.973$ $(95\%\ \mathrm{CI}, 0.901 - 1.045)$, indicating little additional predictive benefit from nonlinear weighting.

We tested the robustness of these results by performing the following analyses: (1) predicting executive control ability derived from three behavioral tasks in an independent UCLA dataset (*n* = 122, **Fig. S6**); (2) predicting specific task performance (**Fig. S7**); (3) constructing the CPM using a nonlinear kernel regression (**Fig. S8**); (4) using a logarithmic transformation, i.e., log (SC+1) (**Fig. S9**); (5) reconstructing SC with an alternative tractography algorithm (**Fig. S10**); (6) increasing network density by setting the desired number of streamlines as 50 million in tractography (**Fig. S10**). These systematic analyses yielded consistent findings, suggesting that nonlinear weighting improves the prediction of general cognitive ability and memory, but not crystallized intelligence or processing speed. Although smaller $\beta$ increased the relative contribution of weak connectivity to the predictions, the connectivity contributions differed across cognitive abilities, exhibiting either graded or threshold-like patterns across the SC weight spectrum (**Fig. S11**).

We examined whether weak connections selected by CPM were topologically organized in terms of topological properties. First, we calculated the weighted degree of regions in the group-averaged SC network, and selected those regions in the top 10% degree as structural hubs (see similar results for functional hubs in **Fig. S12**). Rather than explicitly defining weak connectivity using a threshold, we fixed the proportion of the strongest connections (i.e., $p$=10%~40%) and analyzed the remaining weak connectivity. By defining the number of selected connections in each cross-validation fold in the CPM as $n_{weak}$, and setting the number of connections linking those hubs as $n_{weak}^{hubs}$, we found that the ratio $= n_{weak}^{hubs}/n_{weak}$ for selected connectivity in the CPM was consistently larger than that for unselected connectivity (all $p_{pt}<0.001$, **Fig. 1B**) for different proportions. This enrichment suggests that selected weak connections may provide hub-to-hub parallel pathways relevant to cognitive abilities, consistent with a recent finding [34].

Then, using the optimal $\beta$ identified in the trained models, we performed a series of degree-preserving rewiring analyses to evaluate the contribution of weak connectivity to cognitive predictions, i.e., fixing the proportion of the strongest connections and rewiring the remaining weak connectivity. First, as the proportion of rewired weak connectivity increased, the prediction performance was significantly decreased for general cognitive ability, memory and crystallized intelligence (**Fig. 1C**), suggesting that weak connections contribute to the prediction of cognitive abilities. Second, we respectively rewired the within- and between-module weak connectivity while keeping the equal number of rewired connections. Compared with within-module weak connectivity rewiring, rewiring between-module weak connectivity induced lower prediction performance for general cognitive ability and memory, but higher performance for crystallized intelligence (**Fig. 1D and S13),** consistent with distinct segregation–integration demands across cognitive abilities. Finally, in a test-retest HCP dataset with 44 individuals having two DWI scans, we calculated the intraclass correlation coefficient (ICC) to evaluate the reliability of each connection (see **Supporting Information**). We identified the high/low-reliability weak connections (i.e., ICC>0.6 and ICC<0.4), and then in the primary dataset, we rewired the two types of weak connectivity separately while keeping the equal number of rewired connections. We found that rewiring high-reliability weak connections resulted in lower prediction accuracy than rewiring low-reliability weak connections **(Fig. 1E and S13)**, highlighting the meaningful contribution of highly reliable weak connections to cognitive abilities.

Overall, our results indicate that under nonlinear weighting, weak connections were functionally relevant in predicting cognitive abilities, suggesting that their contribution should not be attributed solely to noise. In fact, although nonlinear weighting may also increase the relative influence of noisy connections, cognitive predictions appeared to be highly nontrivial because the CPM showed substantial robustness to noise, which was explicitly tested by introducing more false positives **(Fig. S14)** and weight-dependent artificial noise to SC (**Fig. S15**). It should also be noted that strong connections still form the structural backbone for cognition, as perturbing equivalent proportions of the strongest connections resulted in larger decreases in prediction performance than perturbing the weakest connections (**Fig. S16**).

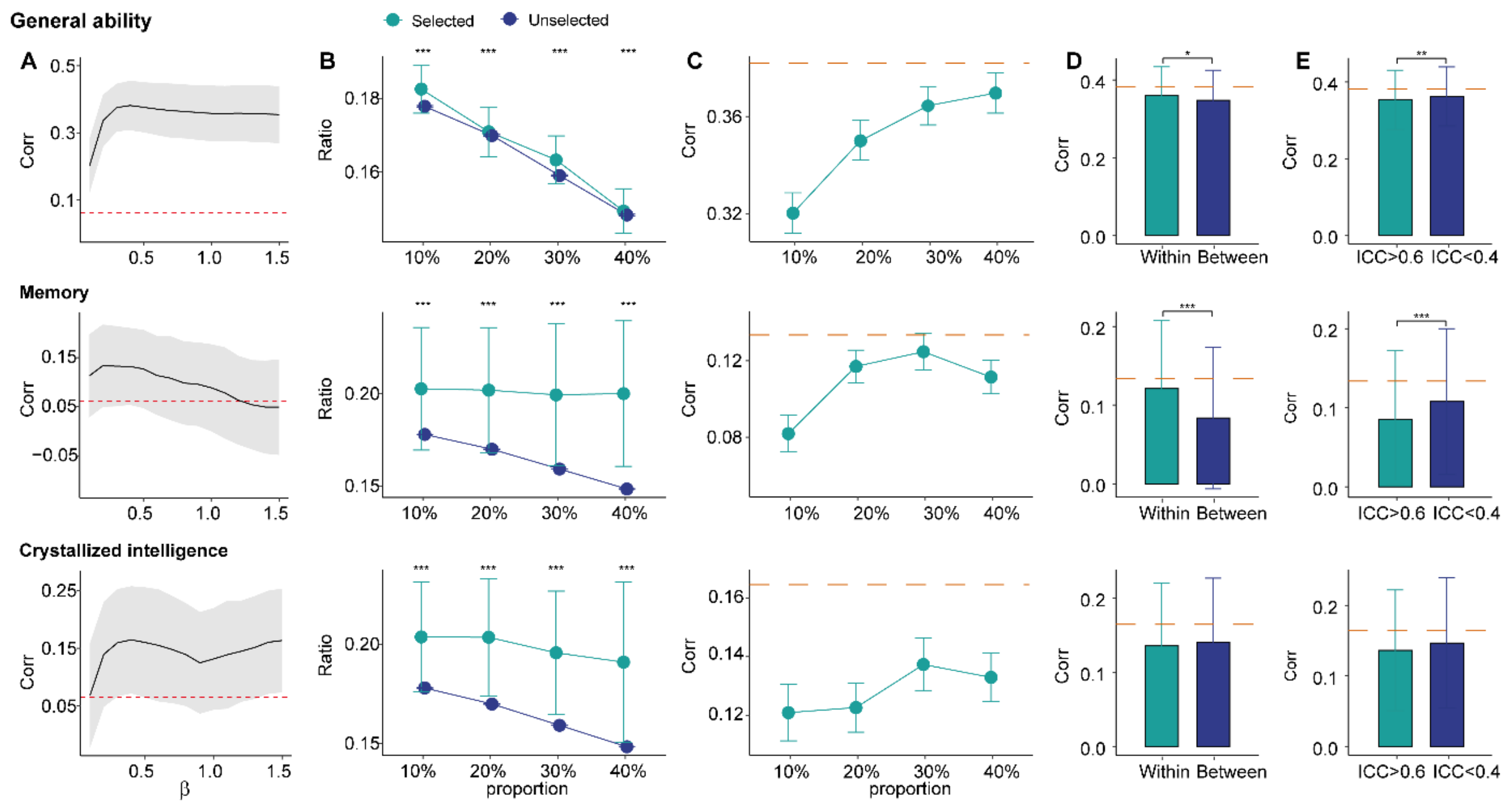

**Fig. 1. Nonlinear weighting improves prediction of general cognitive ability and memory.** **(A)** The trends of mean correlations between predicted and latent scores of cognitive abilities varies with $\beta$. Shaded areas indicate the standard deviation across 100 iterations with training and testing sets randomly assigned. The dashed red line represents the critical correlation threshold (p = 0.05). Here, the $\beta$ was first given and then the prediction was performed, which would induce optimistic bias in selecting optimal $\beta$. So, we designed a nested CPM to calculated the optimal $\beta$ and its 95% confidence interval (CI). **(B)** When fixing the proportion of the strongest connections (i.e., $p$=10%~40%), the ratio of selected versus unselected weak connections linking structural hubs was plotted. Error bar indicates the standard deviation across 100 iterations. **(C)** Prediction of cognitive abilities after rewiring weak connectivity. The red dashed line indicates the best predictions. (**D**) Prediction of cognitive abilities after rewiring within-module or between-module weak connectivity. For general cognitive ability: $t_{pt}(99)=2.268$, $p_{pt}=0.026$, $t_{NB}(99) = 0.652$, $p_{NB} = 0.516$; memory: $t_{pt}(99)=6.594$, $p_{pt}<0.001$; $t_{NB}(99) = 1.895$, $p_{NB} = 0.061$; and crystallized intelligence: $t_{pt}(99)=-0.559$, $p_{pt}=0.578$. **(E)** Prediction of cognitive abilities after rewiring high-reliability (ICC > 0.6) or low-reliability (ICC < 0.4) weak connectivity. For general cognitive ability: $t_{pt}(99)=-3.214$, $p_{pt}=0.002$, $t_{NB}(99) = -0.924$, $p_{NB} = 0.360$; memory: $t_{pt}(99)=-4.133$, $p_{pt}<0.001$, $t_{NB}(99) =-1.188$, $p_{NB} = 0.238$; and crystallized intelligence: $t_{pt}(99)=-0.870$, $p_{pt}=0.387$. Here, the proportion of the strongest connections was 10% (see similar results for other proportions in **Fig. S13**).

**Cognitive prediction gains persist in the reliability-aware connectome**

Because noninvasive tractography is prone to false positives, particularly among weak and long-range connections [14], we next asked whether cognitive prediction gains from nonlinear weighting persist in a more reliable connectome. We applied two post-filtering approaches (SIFT2 [16] and COMMIT2 [3]). SIFT2 mainly decreases the weight of weak connections, but does not remove the links (**Fig. S17**). COMMIT2 introduces biological constraints on the formation of bundles to reduce false-positives [3] while suppressing and removing weak connections to generate different network densities by varying the model parameter $\lambda$ (**Fig. S17**, see Materials and Methods). In SIFT2/COMMIT2 SC, nonlinear weighting also improved the prediction of general cognitive ability and memory, but not for crystallized intelligence **(Fig. S16)**, and SIFT2 SC had similar predictive power as basic SC **(Fig. S16)**. But compared with basic/SIFT2, COMMIT2 produced lower prediction performance for general cognitive ability and memory, not for crystallized intelligence (**Fig. S16**), suggesting that aggressive filtering of likely false-positives streamlines may also attenuate functionally informative connection weights.

Therefore, to reduce likely false-positive contributions and retain informative weights, we constructed a fused SC matrix **(Fig. 2A**) by applying COMMIT2-derived connectivity mask while retaining SIFT2 SC weight. Across different $\lambda$, the predictions of general cognitive ability and memory from fused SC were significantly higher than those from COMMIT2 SC (**Fig. 2B**, **Table S1**), similar to those from SIFT2 SC (**Table S1**). Alternatively, we obtained a "combined SC" (**Fig. 2A**) by retaining all SIFT2 links, using COMMIT2 weights for overlapping connectivity and SIFT2 weights otherwise. This combined SC produced substantially weaker predictions than fused SC (**Fig. 2B**, **Table S1**). Moreover, fused SC had consistently higher ICCs for both high/low-reliability connectivity (ICC>0.6 and ICC<0.4) than COMMIT2 SC in the test-retest dataset (**Fig. 2C**). But stronger biological constraint of bundles in COMMIT2 may discard the putatively plausible connections [3], particularly for off-bundle streamlines [15, 35]. As $\lambda$ increased, the ICC for low-reliability connectivity increased, but a substantial fraction of high-reliability connectivity was also removed (**Fig. 2C**). Prediction performance for general cognitive ability and memory by fused SC initially remained stable and then declined, but not significantly (**Fig. 2B**, **Table S1**). These phenomena were similar when setting the desired number of streamlines as 50 million in tractography (**Fig. 2B**), a procedure expected to increase the prevalence of potential false-positive connections. Thus, the best trade-off between network density and cognitive prediction occurred near $\lambda = 0.05$ (density: 39.6%). In this case, the optimal $\beta$ values were approximately at $\beta$=0.739 (95% CI, 0.680-0.798) for general cognitive ability, $\beta$=0.572 (95% CI, 0.496-0.648) for memory, significantly higher than those for $\beta$=1.0 (g: $t_{pt}(99) = 6.635$, $p_{pt} < 0.001$, $t_{NB}(99) = 1.907$, $p_{NB} = 0.059$; mem: $t_{pt}(99) = 8.044$, $p_{pt} < 0.001$, $t_{NB}(99) = 2.311$, $p_{NB} = 0.023$). The optimal $\beta$ was 1.038 (95% CI, 0.986-1.090) for crystallized intelligence **(Fig. 2B and S16)**. Importantly, these organizational principles were preserved in the reliability-aware connectome (fused SC), where selected weak connectivity remained enriched around hubs, and rewiring high-reliability weak connectivity continued to induce larger reductions in cognitive prediction, and the opposite effects of within- versus between-module rewiring across cognitive domains were also preserved **(Fig. S18)**.

Therefore, even in the reliability-aware connectome, nonlinear weighting robustly improves the prediction of general cognitive ability and memory, supporting the view that weak connections carry biologically meaningful information rather than merely reflecting tractography noise.

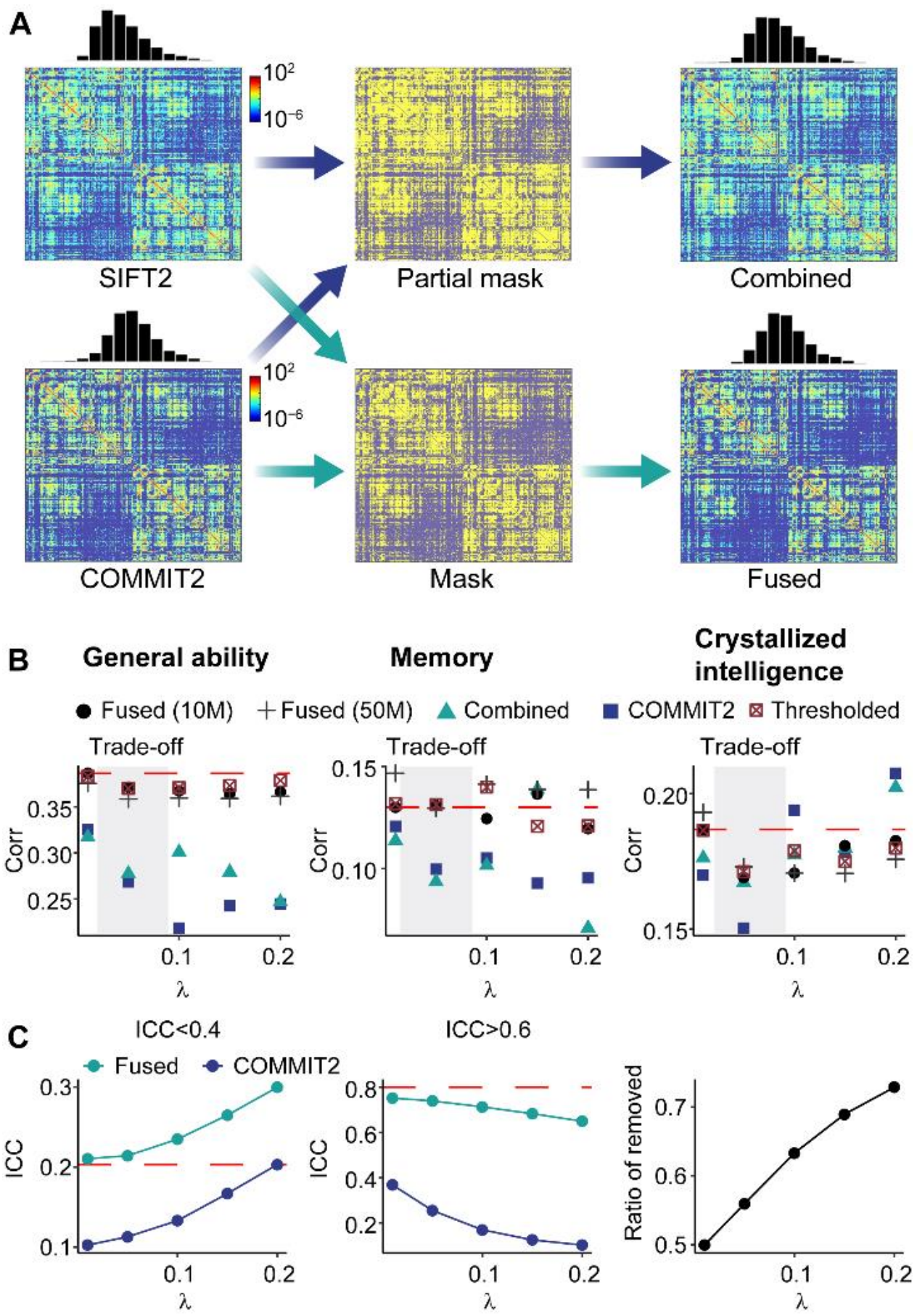

**Fig. 2. A reliability-aware connectome. (A)** Schematic illustration of combined and fused SC derived from SIFT2 and COMMIT2. **(B)** The best predictions of cognitive abilities for fused (settings of 10/50 million streamlines), combined, COMMIT2 and thresholded SC networks with parameter $\lambda$ varying. Error bars were omitted for visual clarity; statistical results were reported in **Table S1**. The red dashed lines indicate the best predictions for SIFT2 SC. The shadow region indicates the trade-off between prediction performance and removal of likely false-positive connections. All analyses were with the 10 million setting except for specific indication. The network density for fused and thresholded connectivity was the same. **(C)** In the test-retest dataset, the ICC for low and high-reliability connectivity (ICC<0.4 and ICC>0.6) in fused SC was varying with $\lambda$, as well as the COMMIT2 SC. The ratio of high-reliability connectivity removed in fused SC was shown in the right panel. Because the fusion was performed at the individual level and ICC was calculated in the group level, we first identified the removed connectivity when this link was absent in 90% of individuals. Then, the ratio of removed link in all high-reliability connectivity was calculated with $\lambda$ increasing.

**Weak connectivity supports brain functional organization under nonlinear weighting**

We next examined whether nonlinear weighting contributes to brain functional organization. In a whole-cortex dynamical model based on fused SC (see Materials and Methods), we optimized the global coupling $c$ and $\beta$ to maximize the similarity between simulated and empirical functional connectivity (i.e., minimizing FC distance), yielding an optimum around $\beta$ =0.2 **(Fig. 3A)**. To test this result while avoiding optimistic bias, we generated two FC matrices by concatenating two fMRI time series in each individual, using one for model fitting and the other for testing. The mean optimal $\beta$ was 0.181 (95% CI, 0.178-0.183,), with the FC distance significantly smaller than that for $\beta$ = 1.0 ($t_{pt}$ (998) = -101.70, $p_{pt} < 0.001$). To ensure that this simulation improvement was not driven by increased model complexity, we calculated the Akaike Information Criterion (AIC) and Bayesian Information Criterion (BIC, see Supplementary Information). The optimal-$\beta$ model had lower AIC ($t_{pt}$ (997)=98.86, $p_{pt}$<0.001) and BIC ($t_{pt}$ (997)=98.86, $p_{pt}$<0.001) than the $\beta$=1.0 model (**Fig. S19**). These results were robust for a nonlinear Jansen-Rit model and a logarithmic transformation, i.e., log (SC+1) (**Figs. S19 and S20**). In addition, the best Pearson correlation between SC and FC in individuals appeared at $\beta$ = 0.319 (95% CI, 0.315-0.322), and the peak correlation was significantly higher than the correlation obtained at $\beta$ = 1.0 ($t_{pt}$ (998) = 151.300, $p_{pt}$ < 0.001, **Fig. 3B**). When using a deep learning model [36] to predict fused SC from FC, the correlation between predicted and empirical SC peaked at $\beta$ = 0.44 (95% CI, 0.32-0.54), significantly higher than that for $\beta$ = 1.0 ($t_{pt}$(99)=11.682, $p_{pt}$<0.001, $t_{NB}$(99)=4.524, $p_{NB}$<0.001, **Fig. 3C**). These results indicate that nonlinear weighting with $\beta$ <1.0 improves the simulation of brain functional connectivity and the structure-function coupling.

We next asked whether the functional gains associated with nonlinear weighting of weak connectivity persist after reliability-aware control. We used the nested spectral partition (NSP) method to quantify segregation-integration balance [32], i.e., $H_B$ = 0. Positive $H_B$ indicates integration bias, whereas negative $H_B$ indicates segregation bias (see Materials and Methods). Resting-state functional networks in healthy young adults maintain a segregation-integration balance [32], as further confirmed by the $H_B$ being close to zero for the empirical and simulated FC (**Fig. S21**). We therefore used the absolute $H_B$ difference between empirical and simulated FC as a measure of large-scale organizational fidelity. As $\lambda$ increased, the absolute $H_B$ difference retained stable until approximately $\lambda \sim 0.05$ and then increased (**Fig. 3D**). Similar results were observed when tractography density was increased to 50 million streamlines, where the absolute $H_B$ difference reached a minimum at $\lambda = 0.01 - 0.05$ (**Fig. 3D**). These findings suggest that moderate removal of unreliable connectivity can improve the functional gains associated with nonlinear weighting, whereas excessive filtering may eliminate functionally informative connections.

Finally, we studied whether weak connections in fused SC contribute to explaining individual difference in cognition and brain function. For each individual, the weakest connections in SIFT2 SC were set to zero until the network density was equal to that of the corresponding fused network at $\lambda = 0.05$. Fused and thresholded SCs yielded similar cognitive-prediction performance for general cognitive ability, memory and crystallized intelligence (**Fig. 2B and Table S1**). However, compared to thresholded SC, fused SC generates smaller FC distance in the model ($t_{pt}$(998) = -44.145, $p_{pt}$ < 0.001, **Fig. 3F**); it also showed stronger empirical SC-FC coupling at the optimal $\beta$ ($t_{pt}$(998) = 21.409, $p_{pt}$<0.001, **Fig. 3E**), but these improvements were not observed at $\beta$ = 1.0 **(Fig. S22)**. The correlation with predicted SC (**Fig. S23A**) and $H_B$ difference (**Fig. S23B**) did not differ significantly difference between fused and thresholded SC.

Therefore, the fusion approach provides a reliability-aware alternative to exclude noise connectivity while retaining weak connections. Thresholding may remove a biologically informative part of the brain structural connectome, and under nonlinear weighting, weak connections contribute incremental gains in brain functional connectivity simulation and structure-function coupling, but do not improve cognitive prediction relative to density-matched thresholding.

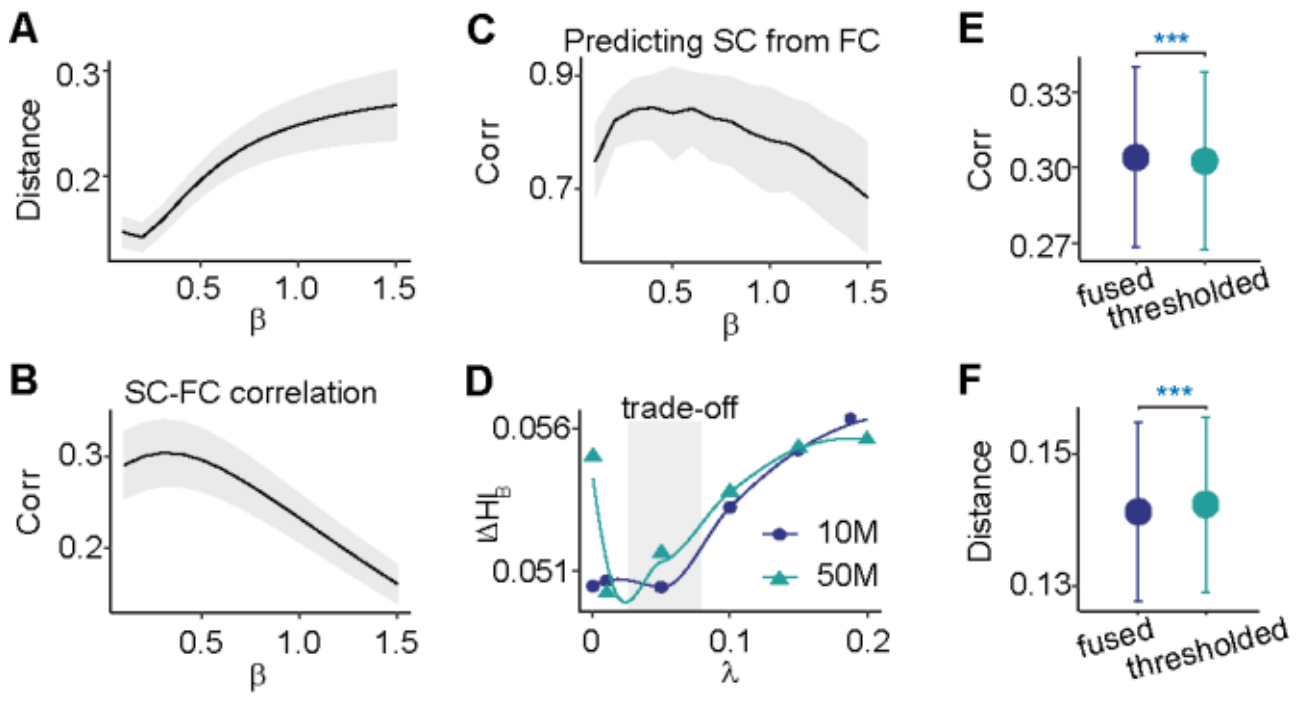

**Fig. 3. Weak connections contribute to the simulation of brain functional organization.** **(A)** The smallest distance between empirical and simulated FC from fused SC ($\lambda = 0.05$) varies with $\beta$. For each $\beta$, the smallest distance across different coupling strengths $c$ was first obtained from each individual and then averaged. The shadow indicates standard deviations across individuals. **(B)** The Pearson correlations between empirical fused SC with different $\beta$ and empirical FC. **(C)** The Pearson correlations between fused and predicted SC from empirical FC using a deep learning model [36]. **(D)** The $H_B$ difference between empirical and simulated FC with varying $\lambda$. For each $\lambda$, the optimal parameters for the smallest distance between simulated and empirical FC were first identified and then the $H_B$ was calculated. **(E)** The comparison of the highest SC-FC correlations for fused and thresholded SC. **(F)** The smallest empirical-simulated FC distance between fused and thresholded SC networks with the same density. All analyses employed the 10 million setting except for (D).

**Weak connections expand segregation–integration repertoire**

We further studied the potential dynamical mechanisms underlying the selective relevance of weak connections for cognitive abilities in terms of segregation and integration. In the large-scale model ($\beta = 0.2$), we retained the proportion of the strongest connections and found that removing weak connections from fused SC reduced $H_B$ (**Fig. 4B**), indicating a disruption of segregation–integration balance and a shift toward higher network segregation. Thus, weak connections contribute to the brain functional balance by primarily enhancing network integration. Consistent with this, weak connections were selectively relevant to general cognitive ability and memory, which are in turn supported by higher integration and the segregation-integration balance [32]; however, crystallized intelligence is thought to rely more on segregated network organization and may therefore benefit less from nonlinear weighting.

Segregation and integration can be further understood through the intrinsic structural eigenmodes of SC networks [37, 38]. The first mode, with an eigenvalue of zero, corresponds to the homogeneous activation of whole brain regions, while higher-order modes describe more localized activation patterns across spatiotemporal scales. We defined the active size in each mode as the number of regions with squared eigenvector component values larger than a threshold (i.e., standard deviation, **Fig. 4A**). As the order of modes increased, the active size decreased (**Fig. 4C**), reflecting that lower-order modes are associated with larger-scale integration and that higher-order modes correspond to more segregated activation of local regions. When weak connections were thresholded (e.g., 10% density), the active size of low-order modes decreased while the size of high-order modes increased (**Fig. 4C**). This phenomenon can be clearly seen in the example of two specific structural modes, wherein the low-order mode in the 10% density network had a smaller spatial extent of activation, and the high-order mode had a larger range (**Fig. 4D**) than in the original fused SC network. Thus, a wide range of regions cannot be coactivated to generate large-scale integration in the low-order mode without including weak connections. However, compared to fused SC network, the high-order modes activate more regions and appear to induce broader local integrations.

By calculating the mean active sizes of low-order and high-order modes, we found that with increasing network density when including weak connections, the active size of low-order modes increased and that of high-order modes decreased (**Fig. 4E**), revealing that weak connections support global integration by increasing the active sizes in low-order modes, while they also contribute to finer-scale segregation in high-order modes. This counterintuitive pattern can be interpreted as follows, as local circuits without the inputs from weak connections projecting from other circuits may become more cohesive due to strong local connections. These local circuits may therefore fail to support finer-scale segregation that may be induced by (weak) projections from other distant regions. Thus, weak connections would greatly expand the modal repertoire of the structural connectome, potentially increasing the diversity of accessible, as supported by the highly heterogeneous active sizes across modes (**Fig. 4F**).

Overall, weak connections support both large-scale integration and fine-scale segregation, thereby expanding the modal repertoire of brain network hierarchy. Under nonlinear weighting, removing weak connections disproportionately alters high-order modes (**Fig. S24**), suggesting a preferential role of weak connections in shaping fine-scale segregation.

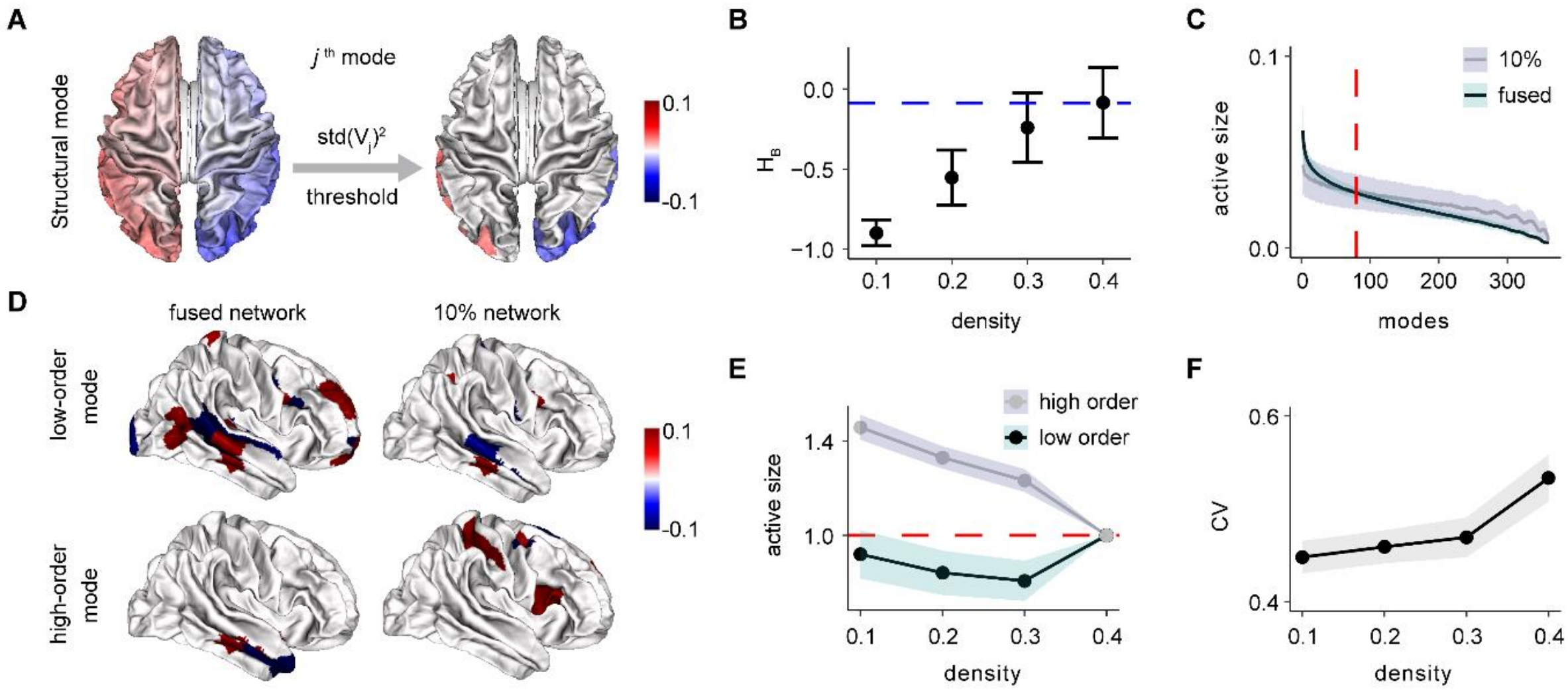

**Fig. 4. Effect of weak connections on segregation and integration.** **(A)** Active size in structural modes was defined as the percentage of significantly active regions (above 1 SD). **(B)** Segregation-integration balance measure $H_B$ varying with network density. The model parameters were fixed as the optimal parameters for fused SC network (**Fig. 3**). The blue dashed line represents the empirical $H_B$ and the error bar indicates the standard deviations across individuals. **(C)** Average active size of structural modes in fused and corresponding 10% thresholded SC networks ($\beta = 0.2$, see similar results for other $\beta$ in **Fig. S24**). The red dashed line separates the modes into low and high orders. **(D)** Brain mappings of two specific examples of corresponding structural modes (7$^{th}$ and 201$^{st}$) for averaged fused and 10% density networks. Regions with eigenvector-component values above the threshold defined in (A) were plotted. **(E)** Mean active sizes (relative to the corresponding size of original fused SC network) of low- and high-order modes for networks with different densities. **(F)** The coefficients of variation (CV) of active size across 2$^{nd}$-N$^{th}$ modes.

**Organizational principles and functional heterogeneity of weak connections.**

We finally investigated how the functional influence of weak connections is related to their organization. By partitioning the brain into seven resting-state networks (RSNs) [39], we calculated the binary densities within and between RSNs in individual SC networks, measured by the number of existing connections divided by the maximum number of possible connections (irrespective of the connection weight). When removing weak connections, both inter- and intra-RSN densities decreased (**Fig. 5A**), but the inter-RSN density decreased more rapidly than the intra-RSN density from fused network to 10% density network (relative change: 77.9% *vs.* 56.6%, $t_{pt}(998) = 558.32$, $p_{pt}<0.001$), indicating that weak connections are distributed between RSNs more than within each RSN (**Fig. S25**), consistent with their contributions to cognitive abilities.

We further investigated the organizational principle of weak connections in terms of gene spatial expression. At the whole-cortex scale, we used Allen Human Brain Atlas gene-expression data to construct a group-averaged gene co-expression (GC) gradient which corresponds to the hierarchy from posterior to anterior cortical regions (**Fig. S26**), consistent with the primary–transmodal gradient from other imaging measures [40-43]. After reordering the SC matrix along the GC gradient, we found that the connections with negative GC spanned longer distances along the GC gradient than connections with positive GC (**Figs. S25 and S26**), and weak connections primarily connect the long-distance gradient locations (**Fig. S25**), i.e., dissimilar brain regions between the posterior and anterior cortex. We then fixed the proportion of the strongest connections and examined the remaining weak connections. We calculated the proportion of weak connections with negative GC among all weak connections incident to each region, and observed a U-shape relationship between the ratio and the GC gradient (the proportion $p$=10%, nonlinear fitting $y \sim x^2 + x$ yielding $R^2 = 0.56$, F(2,357) =232.58, p < 0.001, **Fig. 5B**; permutation test of Moran spectral randomization, p<0.001, **Fig. S27;** see similar results for $p$=20% in **Fig. S28**), wherein visual, motor and limbic systems had higher ratios than other systems. This spatial gradient analysis suggested functional heterogeneity among weak connections. We found that across different proportions ($\beta = 0.2$), deleting weak connections with negative GC induced stronger alterations of active size in structural modes (**Fig. 5D**), larger reduction in modal repertoire (**Fig. 5E**), and stronger deviation from network balance (**Fig. 5F**) than deleting them with positive GC. After excluding a fixed proportion of the strongest connections, we observed that weak connections with negative GC provided superior prediction of general cognitive ability and memory, whereas weak connections with positive GC showed greater predictive power for crystallized intelligence **(Fig. 5C and S29)**.

Overall, these results suggest that weak connectivity with negative GC is more often linked visual, motor, and limbic regions, and has a disproportionately large modeled association with brain network integration, general cognitive ability and memory.

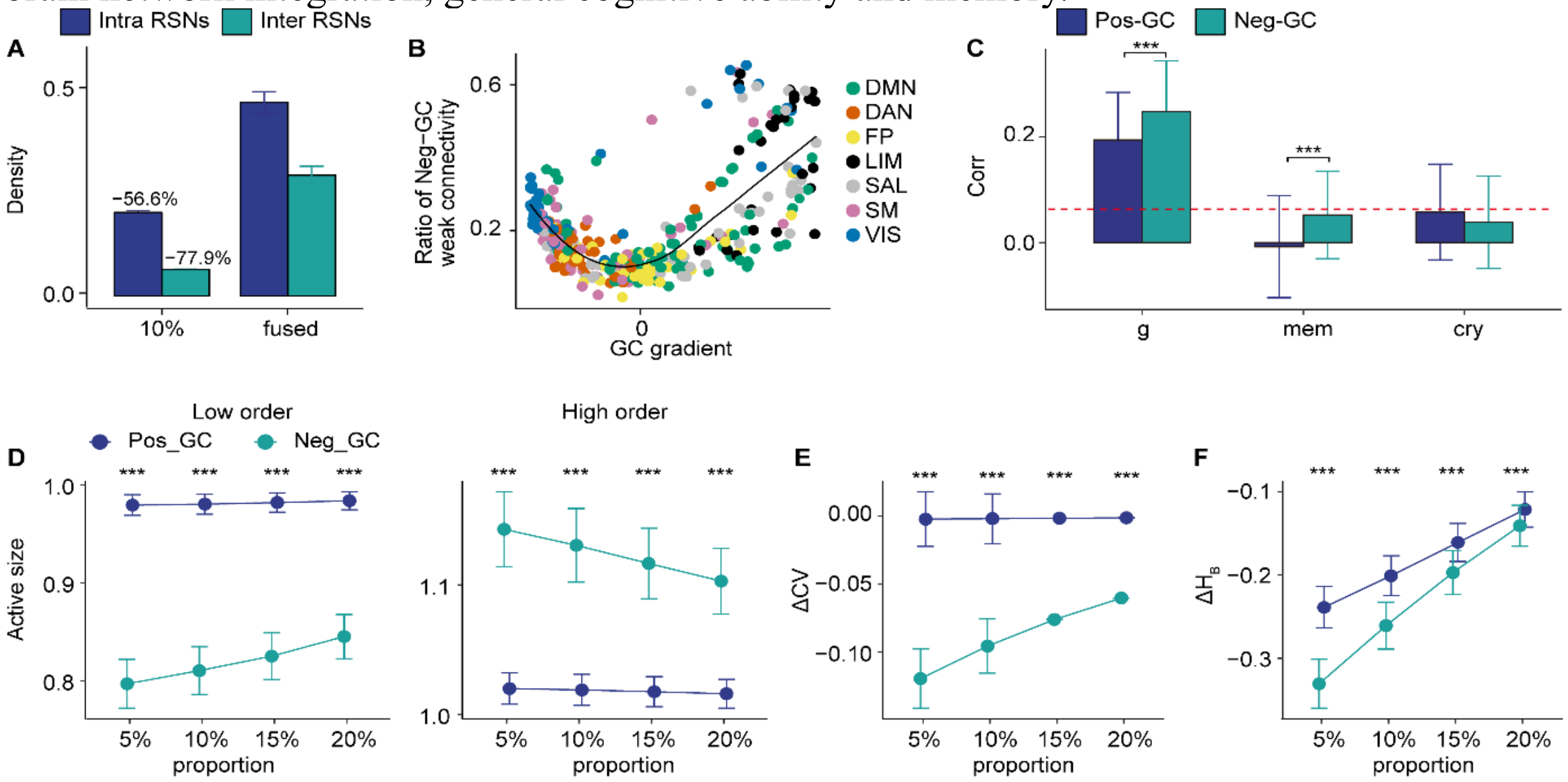

**Fig. 5. Organizational principles of weak connections. (A)** Mean connectivity densities within and between RSNs in individual SC networks. When the network binary density decreased from 39.6% (fused network) to 10%, the relative changes in intra- and inter-RSN densities were shown with the bars. **(B)** The ratio of weak connections with negative GC relative to all connections was calculated for each region and then plotted vs. regional gradient values. The nonlinear fitting line was also plotted. Dorsal attention-DAN; fronto-parietal-FP, default-mode-DMN, visual-VIS, limbic-LIM, somatosensory motor-SM and salience-SAL [39]. (**C**) The predictions of cognitive abilities by weak connections alone with either positive or negative GC under the same splits (the proportion of strongest connection $p$=10%, see similar results for other proportions in **Fig. S29**). To unify the number of input features, the numbers of weak connections with positive and negative GC were controlled to be equal. The dashed red line represents the critical correlation threshold (p = 0.05). For general cognitive ability, $t_{pt}(99)=6.820$, $p_{pt}<0.001$, $t_{NB}(99) = -1.960$, $p_{NB} = 0.053$; memory: $t_{pt}(93)=-4.893$, $p_{pt}<0.001$, $t_{NB}(93)=-1.446$, $p_{NB} = 0.151$; and crystallized intelligence: $t_{pt}(95)=1.678$, $p_{pt}=0.097$. **(D)** Mean active sizes of low-order structural modes and high-order modes with deletion of either weak connections with positive GC (Pos-GC) or that with negative GC (Neg-GC). Mode size was relative to the corresponding size of fused SC network. **(E)** The change of coefficients of variation (CV) of active size across $2^{nd}$-$N^{th}$ modes. **(F)** The change of $H_B$ relative to the fused SC network.

## Discussion

In this work, we addressed the apparent paradox that weak structural connectivity is densely distributed in the human brain but is often considered functionally insignificant and neglected. Across multiple datasets and tractography pipelines, we found that nonlinear weighting consistently enhances the functional relevance of weak connectivity, improving the prediction of general cognitive ability and memory, strengthening structure–function coupling, and improving the fit between simulated and empirical whole-cortex FC. We obtained a reliability-aware connectome that retains weak connections with weights spanning several orders of magnitude and preserves cognitive prediction gains. It also improves structure–function coupling and model-based simulation, outperforming thresholding approaches on both measures. Then, we showed that weak connections not only enhance functional integration and contribute to the segregation-integration balance, but also counterintuitively induce finer segregated brain activation patterns in high-order connectome activation modes, overall expanding the modal capacity of brain networks. Finally, we found that weak connectivity with negative gene-coexpression is more often linked visual/motor and limbic regions and shows disproportionately strong associations with general cognitive ability/memory and brain function. Collectively, our findings suggest that rather than being purely noise, weak connections appear to carry reproducible, functionally informative signal.

### Nonlinear weighting of structural connectivity

Driven by the false-positive problem of non-invasive tractography [3-6], weak connections in the human brain structural connectome, with weight spanning orders of magnitude, have been largely omitted from analysis by thresholding methods [17-23, 44-46]. Although recent studies have cautioned against removing weak connections [17-23], and dense or weak-connection-inclusive connectomes have been used in connectome-based predictive modeling to predict cognitive abilities [47], trait anxiety [48] and hearing ability [49], its functional relevance is still often considered limited [17, 24], particularly under linear weighting schemes in graph measures and dynamical models. More importantly, a group-averaged study reported that weak connections promote complex dynamics by diversifying the inputs and outputs to brain areas [13], suggesting that weak connections may have functional relevance. Although log-scaled forms of SC have been considered [50, 51], there is still a lack of understanding of how weak connections can be functionally significant. To address the question, we proposed a working hypothesis that the functional influence of structural connectivity may follow a nonlinear scaling framework rather than a strictly linear weighting rule. Under this theoretical framework, we found that weak connections contribute to more accurate predictions of cognitive abilities, a stronger relationship of brain structural and functional connectivity and a more realistic simulation of brain functional networks in the whole-cortex dynamic model. These results advance our understanding that weak connections can play functional roles through nonlinear scaling, therefore they should not be treated as uniform noise or removed by default in network analyses, despite their vulnerability to false positives [3-6]. However, it is important to note that $\beta$ reflects how structural connectivity information is utilized within a specific dynamical or predictive framework, and accordingly, the optimal $\beta$ depends on the underlying linear/nonlinear dynamics and the modeling objective. Thus, a universally optimal $\beta$ for the nonlinear transformation may not exist.

Several mechanisms could plausibly contribute to such nonlinear amplification: (1) the weight of weak connections in tractography streamlines is even weaker than the actual fiber projections especially for long-range ones [13, 15]; (2) the functional impact of synapses can be amplified by recurrent excitations and spatial clustering of co-activations [52-54]; For example, in layer 4 in the visual cortex [53], the spatial clustering of co-activations preserved by weak excitatory synapses amplifies the synaptic drives and enhances the contribution of the large number of weak synapses to selective responses of primary visual cortex (V1) neurons to visual stimuli; (3) ample evidence has shown that neural dynamics in local or global scales are organized near critical states which are sensitive to small stimulations [37, 55], thus activation projecting through weak connections may still induce sizable responses. How neural activity amplifies signal transmission through weak long-range white-matter pathways remains to be tested directly in animal experiments, and thus our theory is limited by the lack of direct experimental evidence. But these results suggest that a purely graph-theoretical view is insufficient on its own to understand the complex interactions in the brain connectome, and future modeling studies may help to reveal the nonlinear functional roles of weak connections.

**Reliability-aware connectome**

Weak connectivity is a common feature of mammalian connectomes in human and non-human mammalian brains [13, 27, 29]. However, current tractography methods are limited by false positives [3-6], especially in weak connections. To tackle the false-positive problem and enhance the anatomical plausibility and specificity of tractography, the post-filtering methods such as SIFT2 [16] and COMMIT2 [3] were developed under different tractography-filtering assumptions and hypotheses. In terms of cognitive ability predictions, we found that SIFT2 improves the quality of connection weight and COMMIT2 provides the potential mask of plausible true-positive connections by excluding the noisy connectivity. Crucially, we found that fusing the connection weight of SIFT2 and the connectivity mask of COMMIT2 supports cognitive prediction and FC simulation while reducing likely false-positive connections. Compared to the thresholding method [17-23], this fusing method avoids a purely weight-based threshold and can retain weak connections, especially effective in preserving meaningful individual differences in cognitive performance. More importantly, fused SC contributes to better revealing the brain structure-function coupling and simulating brain functional organization, providing further evidence that weak connections have significant impact on brain function. Therefore, even though there is no ground truth human whole-cortex connectome for directly validating the fused SC, our approaches suggest that combining tractography with complementary filtering strategies yields a more reliable structural connectome for studying large-scale brain networks and inter-individual variability. However, it should be noted that this fusion is performed at the level of connectivity and depends strongly on streamline-level tractography and filtering choices, and fused SC weight should also not be confused with the real fiber strength.

**Weak connections support segregation-integration balance**

We found that weak connections are mainly distributed between functional systems, consistent with previous literature [56]. From a network perspective, weak connections play a role in functional integration and in balancing segregation of systems. Our analysis confirmed this role quantitatively. This may explain why general cognitive ability and memory, which are associated with integration and balance, may be more sensitive to weak connections with the nonlinear functioning manner (i.e., $\beta < 1.0$), but crystallized intelligence that requires higher segregation is insensitive to $\beta$. Meanwhile, maintaining a network segregation-integration balance involves coordinated contributions from both weak and strong connections, which may explain why memory had steeper prediction-$\beta$ curve than general cognitive ability. More importantly, eigenmode analysis revealed that weak connections also contribute to finer-scale segregation in local regions. Thus, including weak connections can allow for the integration of small modules into large modules to support larger-scale integration. Concurrently, weak connections expand the hierarchy of activation modes and allow the generation of finer modules differentially linking with other systems, as reflected by higher coherence in high-order mode levels. Overall, the dynamic range and functional capacity of the brain network can be expanded by the presence of weak connections. These results greatly expand the understanding of the role of weak connections in brain dynamics and ultimately provide a promising dimension for studying personalized connectome in cognitive neuroscience and abnormalities related to brain disorders [8, 57].

**Heterogeneous functions of weak connections**

We found that weak connectivity with negative gene co-expression is mainly linked to visual/motor and limbic systems. Regions in the visual system showed positive gene co-expression with motor and dorsal-attention regions, consistent with the bottom-up salience pathway [58], and it has negative gene co-expression with the limbic and DMN systems, matching the top-down control pathway [58]. Thus, weak intersystem connectivity with positive and negative gene co-expression may motivate future tests of directed or laminar projections. In the human structural connectome, it is still challenging to study forward and backward processing through fiber paths obtained from noninvasive diffusion imaging [59] which cannot provide directionality of projections. Reproducing similar analysis in animal brains may contribute to testing this hypothesis. Meanwhile, based on the primate cortical laminar structure and weighted/directed connectivity data, Mejias and colleagues constructed a large-scale computational model to reveal feedforward-associated gamma oscillations and feedback-related alpha/low-beta oscillations [60]. Thus, the development of a computational neuroscience model with forward and backward projections is promising for understanding the dynamic and functional relevance of weak connections[61]. Additionally, long-range weak connections, primarily believed to transfer excitatory signals, can activate both excitatory and inhibitory neurons [62] and may play potentially inhibitory roles. Thus, weak connections linking different functional systems could have complex implications for information transmission and regulation. Understanding of this biological background should be a priority in network neuroscience and further dynamic model analysis should be conducted to reveal the important roles of weak connections.

**Conclusion**

In summary, our proposed framework suggests that weak connections are functionally relevant to cognitive abilities by integrating different systems to maintain a balance between local segregation and global integration, with the functional influence amplified under nonlinear weighting. Fusing multiple filtering methods is a promising way to select more reliable connections, especially preserving weak connections that contain important information of individual differences. The identified organization provides insights into the specificity of weak connections that mainly links visual/motor to limbic areas and shows a large functional impact. These findings highlight the necessity of considering weak connections for brain functional organization and fill an important knowledge gap about weak connections in the current network neuroscience literature.

## Materials and Methods

***HCP dataset.*** The primary dataset was obtained from the WU-Minn Human Connectome Project (HCP) Young Adult database [63], including data of structural magnetic resonance imaging (MRI), diffusion MRI, resting-state functional MRI (fMRI) and nine behavioral measures collected from 999 healthy young adults (22-35 years old). For each participant, the resting-state fMRI data were recorded for 864 s (1,200 frames) in four high-resolution scanning sessions across two days. Details of the MRI data acquisition are extensively described in [63, 64].

***Diffusion MRI data processing.*** The MRtrix3 software was used to process the DWI data that had undergone the standard HCP minimal preprocessing pipeline [30]. The multi-tissue response functions for white matter (WM), gray matter (GM), and cerebrospinal fluid (CSF) were estimated using the *dhollander* algorithm, followed by multi-shell, multi-tissue constrained spherical deconvolution to compute fiber orientation distributions (FODs) for each tissue type. A five-tissue-type segmentation was generated from the T1-weighted image for anatomically constrained tractography. Whole-cortex probabilistic streamlines were then reconstructed using the iFOD2 algorithm, seeding dynamically based on the WM FOD and producing 10 million streamlines with backtracking enabled (the streamline number of 50 million was also chosen). Then, the brain was partitioned into 360 regions (180 per hemisphere) according to the Multi-Modal Parcellation (MMP) atlas [65]. The resulting seed-to-target maps quantify the percentage of streamlines reaching each target region, but a high likelihood does not necessarily mean a strong white matter fiber [6]. Despite the limitations of the tractography, they are currently the only way to map the structural connectivity of the in vivo human brain. Therefore, we used the average probability of streamlines obtained by averaging streamline probabilities across both seed-target directions to define the weight of connection between two regions, termed structural connectivity (SC). The resulting adjacency matrix $W$ was defined as the brain SC network, where the element represents the connection between regions $i$ and $j$. Notably, tracking weak connections with probabilistic tractography may accumulate more errors, especially for long-range pathways. The key point is that such weak connections have often been treated as functionally irrelevant in network neuroscience[17-23], but may not be merely noise due to technical limitations. Instead, it may reflect true individual differences in cognitive abilities.

At the streamline level, the spherical-deconvolution informed filtering of tractograms (SIFT2) [16] and COMMIT2 [3] were applied to improve the biological plausibility of the resulting tractography, by refining streamlines to better reflect underlying white matter anatomy. SIFT2 was designed to enhance the quantitative accuracy of diffusion MRI tractography [16]. It estimates the FOD within each voxel using constrained spherical deconvolution (CSD) and determines streamline weighting coefficients while reducing the influence of non-white matter pathways in the model. This approach optimizes the streamline weight distributions and allows for more accurate quantification of streamline density while maintaining biological plausibility. COMMIT2 enhances the sensitivity/specificity in tractography by incorporating more prior knowledge about brain anatomy that streamlines represent neuronal fibers and naturally organize into bundles [3]. In this method, variable $y$ represents the diffusion-weighted MRI measurements acquired in the voxels, linear operator matrix $A$ models the water molecule movements, including restrictions inside axons, restrictions outside axons, and isotropic diffusion as seen in cerebrospinal fluid. The target function is:

$$\min_{x \geq 0} \parallel Ax - y \parallel_2^2 + \lambda \sum_{g \in \mathcal{G}} \parallel x^{(g)} \parallel_2 \quad (1)$$

Here, $x$ is the positive weight, encoding the matrix $A$. $\mathcal{G}$ denotes a general partitioning of the streamlines into groups. The coefficient $x^{(g)}$ corresponds to the streamlines in each bundle, and the Lasso parameter $\lambda > 0$ controls the balance between data fidelity and regularization. This additional regularization penalizes the contributions at the group level, promoting solutions that converge to a scheme with a minimal number of bundles, thereby providing a clearer interpretation of the measured diffusion-weighted MRI data. This method can generate different densities of SC networks by varying $\lambda$. In the original algorithm, $\lambda = 0$ corresponds to the classical Commit [66], but here, it indicates the basic tractography. Meanwhile, COMMIT2 is unavailable when setting the desired streamline number as 50 million due to algorithm limitations.

***fMRI data processing.*** fMRI data were preprocessed by the HCP team with the minimal preprocessing pipeline [65]. The pipeline includes spatial artifact removal, surface generation, cross-modal registration and alignment to standard space. More details on these processing steps are described in Ref. [65]. The preprocessed fMRI data were converted to surface space ("CIFTI" format), which consists of 91282 cortical and subcortical grayordinates with a resolution of 2 mm [63]. The blood oxygen level-dependent (BOLD) time series of each of the 360 regions were extracted. Since the global component of the fMRI fluctuations measured during resting state is tightly coupled with the underlying neural activity [67], we did not perform the global signal regression (GSR) [32]. This no-GSR operation contributes to the positive functional connectivity (FC) measured by the Pearson correlation between regional time series, which is necessary to the large-scale dynamic model simulation in the following and brain network segregation-integration balance [32].

***Cognitive abilities.*** To test whether weak connections have functional implications for cognitive abilities, we included cognitive behavioral measures from nine cognitive tasks in the dataset: picture sequence memory (PSM), penn word memory test (PWMT), penn progressive matrices (PPM), variable short penn line orientation test (VSPLOT), picture vocabulary (PV), oral reading recognition (ORR), dimensional change card sort (DCCS), flanker task (FT), and pattern comparison processing speed (PCPS). Then, consistent with our previous analysis [32], we applied a structural equation model (SEM) to extract four cognitive abilities: crystallized intelligence (cry), processing speed (spd), general cognitive ability (g) and memory (mem, see **Fig. S2**). The crystallized intelligence was modeled by including ORR and PV scores, memory ability was based on PWMT and PSM scores, processing speed was based on DCCS, FT and PCPS scores, and the general cognitive ability was based on all task scores.

The SEM analysis was performed using the *lavaan* package in R [68]. The comparative fit index (CFI > 0.95), root mean square error of approximation (RMSEA < 0.08) and standardized root mean-square residual (SRMR < 0.08) were used as criteria of the model [69]. The model fit was acceptable, with CFI=0.974, SRMR=0.030 and RMSEA=0.048.

***Connectome-based prediction model (CPM).*** To predict cognitive abilities from the brain structural connectome, we constructed a connectome-based prediction model (CPM) using the scikit-learn toolbox in Python [33]. The latent score of cognitive ability (e.g., variable y) can be predicted as:

$$y = x_0 + b_1 x_1^{\beta} + b_2 x_2^{\beta} + \cdots + b_m x_m^{\beta} + c\left(\frac{1}{L}\sum_1^L x_i^{\beta}\right) = x_0 + \sum_1^m b_i x_i^{\beta} + c\overline{x^{\beta}} \quad (2)$$

Here, $y$ is a vector for the cognitive ability scores across individuals, $x_i$ is a vector for $i^{th}$ selected connectivity (totally $m$), and $b_i$ is the corresponding learned regression coefficient. The last item is a factor for global connectivity which is the average of all connections (totally L) in brain SC networks and may be associated with cognitive abilities, and $c$ is the corresponding regression coefficient. Controlling this global factor slightly affects the predictions of cognitive abilities in the strict criteria of feature selection (**Fig. S30**). Although a linear regression model was used, the nonlinear kernel regression that implicitly maps the input into a high-dimensional feature space, generated similar results (**Fig. S8**).

Since we hypothesized that weak connections may be more functionally important than what its weak linear weights suggest, we applied an exponent scale parameter $\beta$ for all SC that was not normalized. It is expected that with the nonlinear functioning manner, the model with $\beta < 1.0$ would generate better prediction performances for cognitive abilities than the original null-model with $\beta = 1.0$ that has the linear functioning manner. In this sense, other nonlinear functions would also generate similar results. Thus, we further used the log (SC+1) function to perform the predictions (**Fig. S9**).

To avoid optimistic bias in selecting $\beta$, the full model selection procedure was nested within the training data, while keeping the test data completely unseen until the final evaluation. This model consists of an outer loop and an inner loop. In the outer loop, we employed a 10-fold cross-validation and repeated 10 times, resulting in 100 outer splits. In each split, the data were divided into an outer training set (90%) and an outer test set (10%, the results are similar for 50%:50%, **Fig. S31**). To enforce the family-wise splitting, we utilized a Group K-Fold strategy to ensure that subjects from the same family were assigned to the same fold. In the inner loop, we performed a further 5-fold internal cross-validation within each outer training set. We iterated through the candidate $\beta \in [0.1,1.5]$ with a grid spacing of 0.1, trained the CPM model on the inner training folds, and evaluated them on the inner validation folds. The $\beta$ value that yielded the highest mean correlation across these inner folds was selected as the optimal $\beta^*$ for that specific outer iteration. This optimal $\beta^*$ was used to train the final model on the entire outer training set and the model was subsequently applied to the untouched outer test set to generate the predictions. Within each of the 100 outer splits, we recorded the prediction performance (measured by the Pearson correlation between the predicted and true ability scores) for the optimal $\beta^*$ and compared it with the baseline $\beta$=1.0. This strictly nested approach eliminates optimistic bias by ensuring that the data used to select the optimal $\beta^*$ is entirely independent of the data used to evaluate prediction accuracy. Because the best $\beta^*$ is different at different iterations, we finally reported the mean value of the best $\beta$ and its 95% confidence interval (CI).

In the CPM, no preprocessing that potentially contributes to data leakage was performed, such as data normalization using both datasets. During the feature selection, within each training fold, connections whose weights were significantly correlated with the cognitive score were selected ($p<0.001$, no correction, see similar results for $p<0.01$ and $p<0.05$ with FDR correction in **Fig. S32**). In this process, connections with zero weight were excluded. The selected connectivity was used to train the least absolute shrinkage and selection operator (LASSO) regression model. We used the function LassoCV to select the best regularization parameter through internal cross-validation by maximizing the coefficient of determination, which reflects how much variation in the dependent variable $y$ can be explained by the independent variable $x$. The parameters were set as default (notably, connection weights were not normalized before LASSO) except the total number of iterations was set to 5,000 and the number of folds to 5. This LASSO regression method further suppresses the feature redundancy by setting weight coefficients of unimportant features to be 0 and reduces the complexity of the model, contributing to preventing overfitting. Meanwhile, to test whether the dependence of L1 penalty on connectivity weight affects the results, we standardized the features immediately before they entered the LassoCV function in both training and testing procedures, without applying standardization to any other steps. We found that $\beta$<1.0 still generated significantly better predictions (**Fig. S33**), supporting the effectiveness of nonlinear scaling.

***Large-scale whole-cortex dynamic model.*** The previous study has suggested that the linear model is better in describing the low-fluctuation fMRI dynamics than the nonlinear model [70]. To investigate the functional impact of connections while discarding other influences. we here used a Gaussian linear stochastic model [32, 37, 71, 72]:

$$\dot{x}_i = -x_i + c\sum_{j=1}^{N} w_{ij}x_j + \sqrt{2}\xi_i \quad (3)$$

where $x_i$ represents the neural population activity of the cortical region, $c$ is the global coupling strength, $w_{ij}$ is the SC between regions that was normalized by the maximal eigenvalue of SC matrix. By averaging over the neural states around their mean produced by an ensemble of noise (i.e., $\dot{x}_i \approx 0$) [71], Eq. (3) can be rewritten as $x_i = c\sum_{j=1}^{N} w_{ij}x_j + \sqrt{2}\xi_i$ with the matrix form $(1 - cW)X = \sqrt{2}\xi$ . Then, this model reflects a resolvent-based stochastic propagation across structural paths. According to Ref. [71], we set $Q = (1 - cW)^{-1} \approx e^{cW}$ and obtained $X = \sqrt{2}Q\xi$. The covariance of neural population activities can be expressed as [71]:

$$COV = 2QQ^T \quad (4)$$

The simulated FC matrix can be estimated as:

$$C_{ij} = \frac{COV_{ij}}{\sqrt{COV_{ii}COV_{jj}}} \quad (5)$$

In this model, neural signals propagate through the SC in a linear manner. However, the weights of SC span orders of magnitude; if the response to inputs is defined as a linear superposition, the contribution of neural signals that are transmitted through weak connections to dynamics is negligible, overwhelmed by that through strong connectivity. Thus, a necessary condition for weak connections to be functionally significant is nonlinearity. We speculated that weak connections may be more functionally important in the interregional transmission of neural activation and functional interactions. To investigate a potential mechanism that might support this speculation we also introduced a scaling exponent $\beta$, similar to that used in CPM, to the SC and set $w_{ij} \rightarrow w_{ij}^{\beta}$. Lower $\beta$<1.0 indicates a higher significance of weak connections to brain dynamics relative to strong connections. This scaling parameter accounts for the differences in the orders of magnitude of weak and strong connectivity, but also reduces the dynamic range of connectivity weights.

There are two controlling parameters in the model: $c \in [5{,}400]$ with a grid spacing of 5 and $\beta \in [0.1{,}1.5]$ with a spacing of 0.1. For each pair of $c$ and $\beta$, the simulated FC matrix was obtained and compared with that from resting-state fMRI data. The optimal parameters were obtained when the Euclidean distance between simulated and empirical FC was minimal. We expected that weak connections with $\beta$<1.0 in this nonlinear manner can better recover the real FC at a suitable critical coupling $c$. In this case, the mapping from tractography weight to effective coupling is nonlinear, although the state equation remains linear. To further test that the nonlinear scaling of SC weights contributes to the FC fitting is not constrained by the linearity or nonlinear transform functions, we also used the log(SC+1) function and performed the simulation with a nonlinear Jansen-Rit model[73] (**Fig. S19**).

The HCP dataset provides four fMRI scans across two days in each individual, and we concatenated the time series to generate the individual stable FC that was compared to the simulated FC for a sufficiently long time scale (**Fig. S34**). This fMRI concatenation greatly improved the reliability of FC [32, 74] and ensured that most of the FC weights were positive. We verified that setting the small fraction of negative FC (3.22%) values to zero did not materially affect the model simulation (**Fig. S35**). Because directly comparing the model performance between the best $\beta$ and $\beta$=1.0 constitutes an optimistic bias, we used a cross-validation to test the best parameters ($c$ and $\beta$). We concatenated the two fMRI series at one day and each individual has two FC matrices. One of the FC matrices was randomly selected to fit the model and identify the optimal parameters, and the other was used to test in each individual.

***Network balance between segregation and integration.*** We adopted the nested-spectral partition (NSP) method to measure integration, segregation and their balance in brain FC networks [32]. The FC matrix $C$ can be eigen-decomposed as:

$$C = \sum_{i=1}^{N} \Lambda_i u_i u_i^T, \quad (6)$$

where $\Lambda_i$ and $u_i$ are the eigenvalue and eigenvector of order $i$. The NSP method detects the hierarchical modules in FC networks based on functional modes, ordered by descending order of eigenvalues. The hierarchical modular partition process is:

1. In the 1st mode, all regions have the same positive or negative eigenvector values, reflecting the coactivation of all regions. This model is regarded as the first level with one module (i.e., whole-cortex network).
2. In the 2nd mode, the regions with positive eigenvector values are assigned to a module, and the remaining regions with negative signs form the second module. This mode is regarded as the second level with two modules.
3. According to the positive or negative eigenvector values of regions in the 3rd mode, each module in the second level can be further partitioned into two submodules to form the third level. Successively, the FC network can be partitioned into hierarchical modules of multiple levels for higher-order modes.

At each level, the number of modules $M_i (i = 1, \cdots, N)$, as well as their sizes $m_j (j = 1, \cdots, M_i)$, were recorded. The segregation and integration at each level was defined as [32]:

$$H_i = \frac{\Lambda_i^2 M_i (1 - p_i)}{N} \quad (7)$$

where $p_i = \sum_j |m_j - N/M_i|/N$ is the deviation from the optimized modular size and is used to correct the heterogeneous modular sizes. As the first mode regards the whole-cortex network as a single module and represents global integration, the integration component was computed as:

$$H_{In} = \frac{H_1}{N} \quad (8)$$

and the segregation component was extracted from the remaining modes (2nd to $N$th):

$$H_{Se} = \sum_{i=2}^{N} \frac{H_i}{N} \quad (9)$$

When the segregation component equals the integration component, the FC network has a balanced segregation and integration, as measured by $H_B = H_{In} - H_{Se}$ with a value close to zero.

***fMRI calibration.*** To address the limitation of shorter fMRI series resulting in higher segregation [75], we calibrated the segregation and integration components in individual static FC networks [32]. First, we concatenated all BOLD series from all sessions and individuals to obtain the stable FC across long enough time scale, and the corresponding integration component $H_{In}^S$ and segregation component $H_{Se}^S$ were calculated. The individual integration component was calibrated with $H_{In}^{i\prime} = H_{In}^{i} \frac{H_{In}^S}{\langle H_{In} \rangle}$. Here, $i$ indicates the individual and $\langle \ \rangle$ represents the group average across all individuals. This process was the same for segregation component.

***Predicting structural connectivity using functional connectivity.*** We adopted the default deep learning model [36] to predict the SC using FC. The model combines a generative adversarial network (GAN) and graph convolutional networks (GCNs): multiple GCN layers serve as the generator, and a separate GCN serves as the discriminator. For each individual SC and FC, the generator was trained to create realistic SC matrices and the discriminator differentiates the input SC as real SC (real samples) from the predicted SC (fake samples) generated by the generator (see **Fig. S36**). The connection weight $w_{ij}$ was normalized as [36]:

$$w_{ij} = w_{ij}^{\beta} \text{ and } \boldsymbol{w} = \frac{w - w_{\mu}}{w_{\sigma}} \quad (10)$$

where $w_{\mu}$ and $w_{\sigma}$ are the mean and standard deviation of $\boldsymbol{w}$ across individuals. We adopted the original code from Ref. [36] except that: (1) the data was randomly divided into training and testing sets (90%:10%, see similar results for 50%:50% in **Fig. S37**), (2) an inner loop was added in the training set to determine the best $\beta$ before testing, and (3) the twins information was considered to enforce the family-wise splitting.

***Structural connectome modes***. As brain structural modes form the orthonormal basis, cortical activity can be expressed in terms of the combinations of structural modes [37, 38], making them a powerful tool to study the effect of weak connections on functional activation patterns. For the SC network with scaled weight, i.e., $w_{ij}^{\beta}$, the Laplacian matrix was first constructed and then the structural modes were obtained through:

$$L = \sum_{k=1}^{N} \lambda_k V_k V_k^T, \quad (11)$$

where $\lambda_k$ and $V_k$ are the eigenvalue and eigenvector of order $k$. These eigenmodes can be understood as an extension of the Fourier basis to the cortical space describing basic patterns of activity at different temporal frequencies. Since eigenvalues correspond to temporal frequencies related to the time activity of the associated modes [37, 38], we ranked the structural modes in an ascending order according to their eigenvalues.

***Alignment of structural modes.*** We expected that the removal of weak connections has a localized effect on several modes and the pattern of other modes is maximally maintained, such that the effect of weak connections on functional activity can be precisely studied. However, in reality, the modes with the maximally similar pattern before and after removing the weak connections would have different orders due to their slightly different eigenvalues, making the direct comparison of modes challenging. To solve this issue, the structural modes from the full and thresholded SC networks should be aligned first. We set the individual full network modes as references and minimized the angle variance of thresholded network modes to the reference modes. Specifically, assuming $U$ to be the eigenvectors of the Laplace matrix of a thresholded SC network, we can calculate the angle $\theta_{ij}$ between the eigenvector $U_i$ and the reference eigenvector $V_j$ of the Laplace matrix of the full SC network. This angle was defined as:

$$\theta_{ij} = \cos^{-1}(U_i \cdot V_j), (i,j = 1,2,\dots,N) \quad (12)$$

Solutions to this assignment problem consist of finding a bijection $f: U \to V$ such that the cost function $\sum_{i,j} \theta_{ij}$ is minimized. A solution can be written as a matrix $X$ where $X_{ij} = 1$ if the eigenvector $U_i$ was assigned to the reference $V_j$, and 0 otherwise. Moreover, $\sum_{i=1}^{N} X_{ij} = 1, \forall j \in \overline{1,N}$ and $\sum_{j=1}^{N} X_{ij} = 1, \forall i \in \overline{1,N}$, so that each vector in $U$ is uniquely aligned with one reference vector in $V$. The Kuhn-Munkres algorithm finds $X$ such that $\sum_{i=1}^{N} \sum_{j=1}^{N} X_{ij} \theta_{i,j}^{k,ref} \to min$ in $O(N^3)$ time [76]. This alignment was performed for each individual. However, notably, this algorithm minimizes the total angle variance but does not guarantee the best alignment between two modes.

***Active size of structural modes.*** Brain structural modes correspond to the hierarchical modules in brain SC networks, and high-order modes correspond with small module sizes to support more localized segregation [37]. To reflect this, we calculated the active size of the structural modes. For the $j^{\text{th}}$ order structural mode $V_j$, its squared form is $v_j = V_j^T V_j$ with $\sum_{i=1}^{N} v_{ji} \equiv 1$ across all regions, and the number of highly active regions was defined as:

$$S_j = \frac{n_{v_j > th}}{N} \quad (13)$$

Here, $n_{v_j > th}$ is the number of regions with $v_j$ larger than the threshold *th*. Since the fluctuations (variance) of structural modes increase with their orders (**Fig. S38**), we set *th* to the standard deviation of $v_j$ across all modes of the reference matrix. This threshold is higher in higher-order modes with larger eigenvalues (frequencies), reflecting the more intense activation of local regions. After the modes from thresholded networks were aligned to the reference modes from original full individual SC networks, each mode's order for different network densities has the same *th* as the reference mode. Meanwhile, the reference structural modes were sorted in descending order of active sizes, and the aligned modes were consequently reordered. Then, the normalized active sizes for low-order modes and high-order mode were separately calculated:

$$S_{low} = \frac{\sum_{j=2}^{180} S_j}{\sum_{j=2}^{180} S_j^{ref}} \text{ and } S_{high} = \frac{\sum_{j=181}^{N} S_j}{\sum_{j=181}^{N} S_j^{ref}} \quad (14)$$

Here, $S_j^{ref}$ refers to the active sizes of the reference modes from original full SC networks. Since the 1st mode with $V_1 = \frac{1}{\sqrt{N}}(1,1,\ldots,1)^T$ and $\lambda_1 = 0$ represents a homogeneous state with all regions equally involved, it is not considered here.

***Gene-coexpression gradient.*** The gene expression data were obtained from the Allen Human Brain Atlas (AHBA) [77]. This atlas contains ~3,700 tissue samples from six donors, and their Montreal Neurological Institute (MNI) coordinates were provided. Tissue samples from four donors covered the left hemisphere, and samples from the remaining two donors covered the whole brain. Since our analysis is on the whole-cortex scale, the gene expression data in both the left and the right hemispheres averaged across all donors were used. Arnatkeviciute et al. developed the *abagen* toolbox to process AHBA genetic data [78]. Here, we used the default parameter settings except for 1) mirroring microarray expression samples in the left hemisphere to the right hemisphere to increase spatial coverage (e.g., parameter *lr_mirror*='leftright'); and 2) filling the empty expression values of the region by interpolating the nearest samples (e.g., parameter *missing*='interpolate'). The results are robust to these two parameters (**Fig. S39**). The final microarray gene expression data were mapped to 360 regions in the MMP atlas. In these preprocessed data, each region contains 15677 genes. The Pearson correlation between the gene expression profiles of regions *i* and *j* was calculated to represent the gene coexpression (GC) between these two regions, and a 360×360 gene coexpression network was constructed.

To calculate the hierarchical gradients, the GC matrices were first transformed into cosine distance matrices, and then the classical multidimensional scaling (CMS) method was applied to obtain the coordinates of regions in the low-dimensional space using the *cmdscale* function in Matlab, which were regarded as the brain gradients. Here, we selected the first components as principal gradients for GC (with 85.3% of variance explained).

***Statistical analysis.*** The paired t-test was used, and the t-statistic and the degree of freedom (df) were provided, i.e., t(df). The Nadeau-Bengio-type correction was further performed to deal with the dependent observations. In Fig. 5B, nonlinear/linear regression models were applied, and the permutation test was performed to further confirm the results. The GC network was first randomly rewired to obtain the null model, and the nonlinear/linear regression model was used to compare the $R^2$ (coefficient of determination) between real and rewired networks. The statistical *p*-value was obtained across 1000 permutations. The permutation test of Moran spectral randomization was based on BrainSpace toolbox.

## Acknowledgments

This work was partially supported by the STI 2030-Major Projects (2022ZD0208500), the National Natural Science Foundation of China (Nos. 12132012, 32541016, 12272292, 11975194), the Hong Kong Research Grant Council (grants SRFS2324-2S05, GRF 12202124, GRF12201421, GRF12200620 and GRF12301019), the Hong Kong Baptist University Research Committee (IG-FNRA/20-21/SCI04, RC_SFCRG_23_24_SCI_06), the German Research Foundation (DFG) to Andrea Hildebrandt (HI 1780/7-1) and Carsten Gießing (GI 682/5-1) as part of the DFG priority program "META-REP: A Meta-scientific Programme to Analyse and Optimise Replicability in the Behavioural, Social, and Cognitive Sciences" (SPP 2317). This research was conducted using the resources of the High Performance Computing Cluster Centre, Hong Kong Baptist University, which receives funding from RGC, University Grant Committee of the HKSAR and HKBU, and the High Performance Computing Platform of XJTU.

## Competing interests

Authors declare that they have no competing interests.

## Data Availability Statement

The MRI and cognitive datasets are available at http://www.humanconnectome.org/study/hcp-young-adult. AHBA gene data are available at http://human.brain-map.org/. The code used for this study is available at https://github.com/TobousRong/weak-connectivity. The deep learning code to predict SC from FC is available at https://github.com/qidianzl/Recovering-Brain-Structure-Network-Using-Functional-Connectivity.